\documentclass[12pt,preprint]{aastex}

\begin{document}

\title{Three-Dimensional Statistics of Radio Polarimetry}
\author{Mark M. McKinnon}
\affil{National Radio Astronomy Observatory\altaffiltext{1}{The National
Radio Astronomy Observatory is a facility of the National Science
Foundation operated under cooperative agreement by Associated 
Universities, Inc.}, Socorro, NM \ \ 87801\ \ USA}

\begin{abstract}

  The measurement of radio polarization may be regarded as a three-dimensional 
statistical problem because the large photon densities at radio wavelengths 
allow the simultaneous detection of the three Stokes parameters which completely 
describe the radiation's polarization. The statistical nature of the problem 
arises from the fluctuating instrumental noise, and possibly from fluctuations 
in the radiation's polarization. A statistical model is used to derive the 
general, three-dimensional statistics that govern radio polarization measurements. 
The statistics are derived for specific cases of source-intrinsic polarization,
with an emphasis on the orthogonal polarization modes in pulsar radio emission. 
The statistics are similar to those commonly found in other fields of the physical,
biological, and Earth sciences. Given the highly variable linear and circular 
polarization of pulsar radio emission, an understanding of the three-dimensional 
statistics may be an essential prequisite to a thorough interpretation of pulsar 
polarization data.

\end{abstract}


\section{INTRODUCTION}

  The large photon densities at radio wavelengths, unlike the much smaller 
densities at much shorter wavelengths, allow the simultaneous detection of the 
three Stokes parameters which completely describe the polarization of the 
electromagnetic radiation (Radhakrishnan 1999; Thompson, Moran, \& Swenson 2001). 
The randomly fluctuating instrumental noise introduces a statistical element to 
the measurements, so a complete and proper interpretation of radio polarization 
observations requires a three-dimensional statistical analysis. Interestingly, 
but quite understandably, radio astronomers do not fully exploit this unique 
aspect of radio science, and effectively reduce a polarization measurement to 
a problem of one or two dimensions by investigating the circular or linear 
polarization, respectively, on an individual or separate basis. 

  The polarization signature of the radiation mechanism under investigation 
usually dictates which Stokes parameters are actually measured. For example, 
the Zeeman effect produces a distinctive, circularly-polarized signal, and the 
measurement of a single component of the polarization vector, the Stokes parameter 
V, is usually sufficient to observe the polarization of the effect. Synchrotron 
radiation produces significant linear polarization, but no circular polarization 
(e.g. Rybicki \& Lightman 1979), so most polarization observations of 
synchrotron-emitting sources only utilize the Stokes parameters Q and U. It follows 
that the measurement of circular polarization is a one-dimensional (1D) statistical 
problem because only one Stokes parameter is involved. Similarly, the measurement 
of linear polarization is a two-dimensional (2D) statistical problem because two 
Stokes parameters are used. The statistics applicable to the measurement of 
circular and linear polarization are well-understood, and are documented in 
McKinnon (2002).

  Three-dimensional (3D) statistics of radio polarimetry are particularly relevant
to pulsar radio emission, where both the linear and circular polarization can be 
significant and highly variable. The radio emission mechanism of pulsars is not 
completely understood, and single pulse polarization observations are made in an 
attempt to understand it. The results of these observations are usually displayed 
as longitude-dependent histograms of fractional linear polarization, fractional 
circular polarization, and position angle (e.g. Backer \& Rankin 1980; Stinebring 
et al. 1984) or as joint probability densities of linear polarization and position 
angle (PA) or of circular polarization and PA (e.g. Cordes, Rankin, \& Backer 1978). 
The histograms and probability densities beautifully illustrate the statistical 
nature of the emission's polarization fluctuations and the fascinating orthogonal 
polarization modes (OPM). However, these display methods, by their very nature, 
force one to interpret the emission's linear and circular polarization on a 
separate basis (e.g McKinnon 2002), and reveal little, if anything, about the 
association of the two polarizations. The display methods, in effect, betray our 
reliance upon the familiar, but limited, 1D and 2D polarization statistics. 
Applying the more general, and more descriptive, 3D statistics to pulsars may be 
essential to understanding their emission mechanism. One reason why the 3D 
statistics may not have been applied is it appears they have yet to be determined.

  Directional statistics are a subset of the more general 2D and 3D vector
statistics, and govern the possible orientations of a vector. They are particularly 
applicable to experiments where a vector's orientation, but not its amplitude, can 
be measured, such as measurements of the vanishing angles of birds taking flight 
and measurements of magnetic remanence in rock formations (e.g. Mardia 1972). The 
statistics' widespread applicability in biology and the Earth sciences, the 
pioneering statistical work of R. A. Fisher, and the advent of the modern computer 
have all contributed to the rapid development of directional statistics over the 
last 30 years (Fisher, Lewis, \& Embleton 1987). Interestingly, directional 
statistics were originally developed to solve a problem in astronomy. In 1734, 
Bernoulli used the statistics to prove that the orbital planes of the planets in 
the Solar System could not be aligned by chance (Mardia 1972, and references therein). 
The statistics are not widely used in radio astronomy now, but the possibility of 
their application to radio polarimetry is very intriguing.

  The purpose of this paper is to derive and summarize the 3D statistics that are 
applicable to radio polarimetry. In \S\ref{sec:fixed}, the statistics of the 
measured polarization vector are derived for a source of fixed polarization using 
the statistical model described in McKinnon (2002). The specific cases of no 
polarization, linear polarization only, and circular polarization only are 
evaluated. The polarization vector statistics for OPM are derived in 
\S\ref{sec:opm}. The directional statistics of a polarization vector are derived 
in \S\ref{sec:direct} for the cases of fixed polarization and OPM. The 
applicability and limitations of the statistics are discussed in 
\S\ref{sec:discuss}. The 3D statistics will be applied to polarization observations
in a future paper.

\section{FIXED POLARIZATION}
\label{sec:fixed}

  The polarization of any source of electromagnetic radiation is completely 
described by the Stokes parameters Q, U, and V. The parameter V describes the 
circular polarization of the radiation, and the parameters Q and U describe the 
real and imaginary parts, respectively, of its linear polarization. The Stokes
parameters form a right hand, Cartesian coordinate system, and the radiation's
polarization can be represented by a vector in this three dimensional space,
which shall be designated as \lq\lq Poincar\'e space" in what follows. For the 
general case of a polarization vector with an amplitude $p$ and an orientation 
specified by an azimuth $\phi_o$ and a colatitude $\theta_o$, the circular 
polarization is $\mu_V = p\cos\theta_o$, and the real and imaginary parts of the 
linear polarization are $\mu_Q = p\sin\theta_o\cos\phi_o$ and 
$\mu_U = p\sin\theta_o\sin\phi_o$, respectively. 

  The Stokes parameters that are actually recorded during an observation include 
the contributions from the Gaussian instrumental noise of the radio telescope.
If $X_N$ is a random variable that represents the instrumental noise, the 
measured Stokes parameters are (McKinnon 2002)

\begin{equation}
\label{eqn:Qfix}
{\rm Q} = \mu_Q +  X_{\rm N,Q},
\end{equation}
\begin{equation}
\label{eqn:Ufix}
{\rm U} = \mu_U + X_{\rm N,U},
\end{equation}
\begin{equation}
\label{eqn:Vfix}
{\rm V} = \mu_V + X_{\rm N,V}.
\end{equation}
Thus defined, each of the measured Stokes parameters is a Gaussian random variable 
with mean $\mu$ and standard deviation $\sigma_N$, where $\sigma_N$ is the 
magnitude of the instrumental noise. Equations~\ref{eqn:Qfix} through~\ref{eqn:Vfix} 
are a set of linear equations that form a simple, statistical model for the 
polarization measured with a radio telescope. In its simplest form, the model 
completely ignores any systematic effects such as instrumental imperfections, 
calibration errors, parallactic angle rotation, interstellar scintillation, and 
Faraday rotation.

  As long as the antenna temperature produced by the radio source is small in 
comparison to the system temperature of the radio telescope, the instrumental noise 
in each of the measured Stokes parameters is independent from that in the others.
This being the case, the measured Stokes parameters are independent of one another,
and their joint probability density is the product of their individual probability 
densities (i.e. $f_{QUV} =f_Qf_Uf_V$). The Stokes parameters can be evaluated in the 
context of a joint probability density because they can be detected simultaneously
at radio wavelengths and thus the polarization state of the radiation can be 
completely determined at any instant.

  The possible amplitudes and orientations of the measured polarization vector can 
be determined by converting the joint probability density from Cartesian to spherical 
coordinates. If the random variables $x$, $y$, and $z$ represent the measured Stokes 
parameters Q, U, and V, respectively, the measured amplitude of the polarization 
vector is $r=(x^2+y^2+z^2)^{1/2}$, and its measured orientation\footnote{This 
nomenclature for orientation angles is consistent with that of traditional, spherical 
coordinate systems. It is different from the nomenclature used in McKinnon \&
Stinebring (1998) and McKinnon (2002) where azimuth and PA were denoted by $\theta$ 
and $\phi$, respectively.} is given by a colatitude of $\theta = \arccos(z/r)$ and 
an azimuth of $\phi=\arctan(y/x)$. The position angle (PA) of the linear polarization 
is $\psi = \phi/2$. After converting from Cartesian to spherical coordinates, one can 
show that the joint probability density governing the measured values of the 
polarization vector's amplitude and orientation is

\begin{equation}
f(r,\theta,\phi) = {r^2\over{\sigma_N^3}}{\sin\theta\over{(2\pi)^{3/2}}}
                   \exp\Biggl[-{(r^2+p^2)\over{2\sigma_N^2}}\Biggr]
                   \exp\Biggl\{{rp\over {\sigma_N^2}}[\cos\theta_o\cos\theta
                 + \sin\theta_o\sin\theta\cos(\phi-\phi_o)]\Biggr\}.
\label{eqn:jd}
\end{equation}

\subsection{No Polarization}

  If the electromagnetic radiation is not polarized, the amplitude of its intrinsic 
polarization vector is $p=0$.  From equation~\ref{eqn:jd}, the joint probability 
density for the measured polarization vector is

\begin{equation}
f(r,\theta,\phi) = {r^2\over{\sigma_N^3}}{\sin\theta\over{(2\pi)^{3/2}}}
                   \exp\Biggl(-{r^2\over {2\sigma_N^2}}\Biggr).
\end{equation}
The distribution of the vector's amplitude is found by integrating the joint 
probability density over all orientation angles.
\begin{equation}
f(r) = \int_{0}^{2\pi}\int_{0}^{\pi}f(r,\theta, \phi)d\theta d\phi =
       {r^2\over{\sigma_N^3}}\sqrt{{2\over{\pi}}}
       \exp\Biggl(-{r^2\over {2\sigma_N^2}}\Biggr)
\label{eqn:MB}
\end{equation}

\noindent The amplitude distribution is identical in functional form to the 
Maxwell-Boltzmann distribution for the speed of molecules in an ideal, classical 
gas. In retrospect, this is not surprsing considering that similar assumptions, 
namely the statistical independence of the Stokes parameters and of the vector 
components of a molecule's velocity, were made in the derivation. The mean, 
$\langle r\rangle$, and standard deviation, $\sigma$, of the amplitude distribution 
are
\begin{equation}
\langle r\rangle = 4\sigma_N/\sqrt{2\pi},
\end{equation}
\begin{equation}
\sigma = \sigma_N\sqrt{3-8/\pi}.
\end{equation}
The distribution of the vector's orientation is
\begin{equation}
f(\theta,\phi) = \int_0^\infty f(r,\theta,\phi)dr = {1\over {4\pi}}\sin\theta,
\label{eqn:iso}
\end{equation}
which is isotropic.

\subsection{Linear Polarization}

  When the radiation is linearly polarized, the colatitude of the intrinsic 
polarization vector is $\theta_o = \pi/2$. If the vector azimuth is $\phi_o = 0$, 
the polarization occurs in the Stokes parameter Q, and the intrinsic Stokes 
parameters are $\mu_Q=p$ and $\mu_U=\mu_V=0$. The joint probability density for 
the measured values of this polarization vector's amplitude and orientation is

\begin{equation}
f(r,\theta,\phi) = {r^2\over{\sigma_N^3}}{\sin\theta\over{(2\pi)^{3/2}}}
                   \exp\Biggl[-{(r^2+p^2)\over{2\sigma_N^2}}\Biggr]
                   \exp\Biggl({rp\sin\theta\cos\phi\over {\sigma_N^2}}\Biggr).
\label{eqn:jdl}
\end{equation}

\noindent The distribution of the vector's amplitude is

\begin{equation}
f(r) = \int_{0}^{2\pi}\int_{0}^{\pi} f(r,\theta,\phi)d\theta d\phi = 
       {r\over{p\sigma_N}}\sqrt{{2\over{\pi}}}
       \exp\Biggl[-{(r^2+p^2)\over {2\sigma_N^2}}\Biggr]
       \sinh\Biggl({rp\over{\sigma_N^2}}\Biggr).
\label{eqn:amp}
\end{equation}

\begin{figure}
\plotone{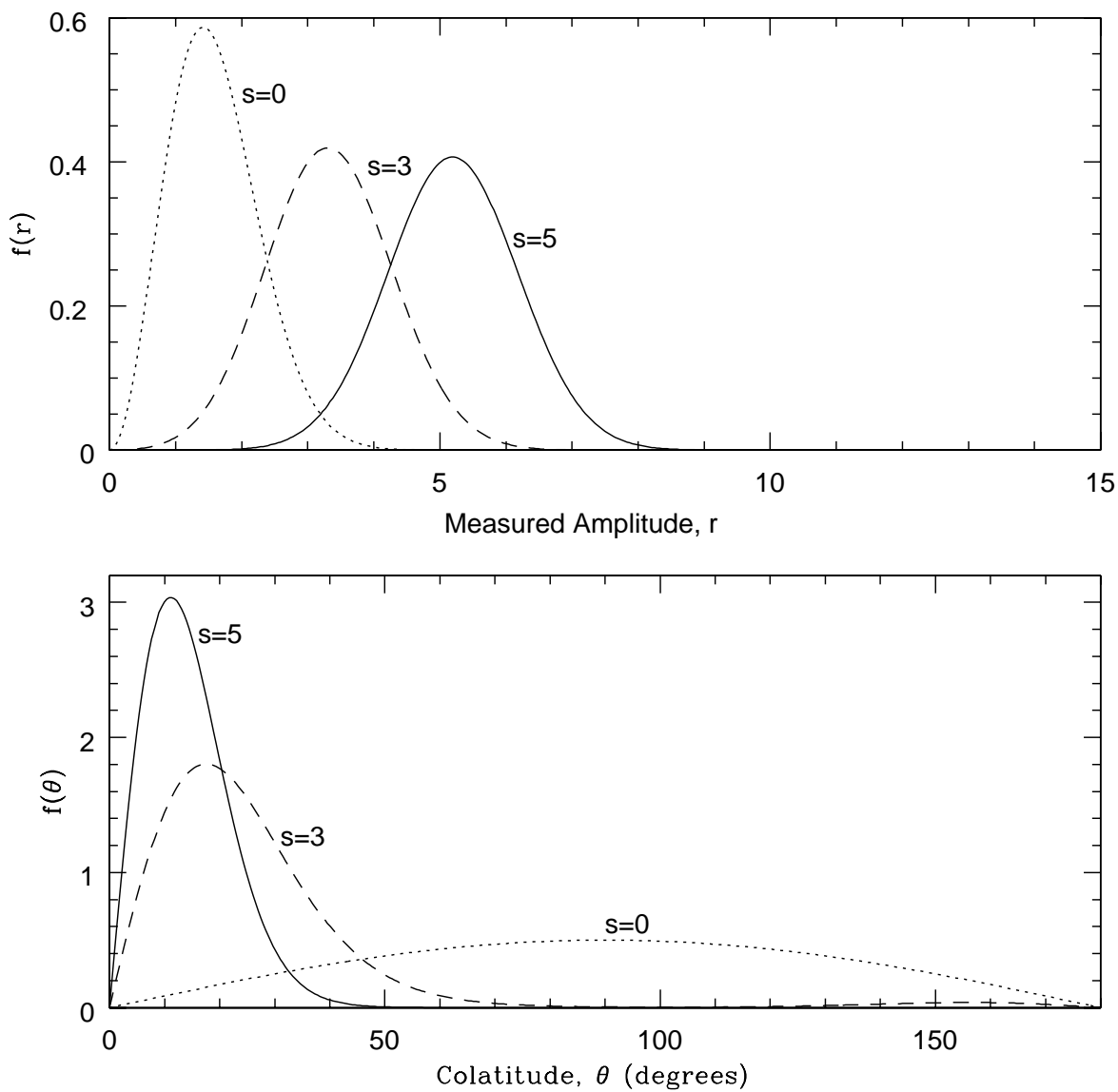}
\caption{Distribution of a polarization vector's measured amplitude (top panel)
and colatitude (bottom panel) for radiation with fixed circular polarization. 
The distributions are shown for different signal-to-noise ratios, $s=p/\sigma_N$. 
The colatitude distribution evolves from an isotropic distribution at $s=0$ to
a Rayleigh distribution for large $s$. The value of the instrumental noise used
in the amplitude distributions is $\sigma_N=1$.}
\label{fig:fixed}
\end{figure}

\noindent The amplitude distribution is shown in the top panel of 
Figure~\ref{fig:fixed} for different signal-to-noise ratios ($s=p/\sigma_N$). As 
can be seen in the figure, the amplitude distribution is Maxwell-Boltzmann for small 
$s$, and resembles a Gaussian when $s$ is large. The mean amplitude is
\begin{equation}
\langle r\rangle = \int_{0}^{\infty} rf(r)dr = 
     \sigma_N\sqrt{{2\over{\pi}}}\exp\Biggl(-{p^2\over {2\sigma_N^2}}\Biggr)
    + \Biggl({p^2 + \sigma_N^2\over {p}}\Biggr){\rm erf}\Biggl({p\over 
      {\sigma_N\sqrt{2}}}\Biggr).
\label{eqn:avgamp}
\end{equation}

\noindent In the limiting cases of high and low signal-to-noise ratio, $s$, the 
mean amplitudes are
\begin{equation}
\langle r\rangle \simeq p + \sigma_N^2/p, \qquad p >> \sigma_N, 
\end{equation}

\begin{equation}
\langle r\rangle \simeq \sigma_N\sqrt{{2\over{\pi}}}\Biggl(2 +
                      {s^2\over{3}}\Biggr), \qquad p < \sigma_N.
\end{equation}

\subsection{Circular Polarization}

  The colatitude of the radiation's polarization vector is $\theta_o = 0$ when 
it is circularly polarized. The Stokes parameters of the radiation's polarization 
are $\mu_V=p$ and $\mu_Q=\mu_U=0$. The joint probability density for the measured 
values of this polarization vector's amplitude and orientation is

\begin{equation}
f(r,\theta,\phi) = {r^2\over{\sigma_N^3}}{\sin\theta\over{(2\pi)^{3/2}}}
                   \exp\Biggl[-{(r^2+p^2)\over{2\sigma_N^2}}\Biggr]
                   \exp\Biggl({rp\cos\theta\over {\sigma_N^2}}\Biggr).
\label{eqn:jdc}
\end{equation}

\noindent By integrating over all orientation angles, one can show that the 
distribution of the vector's amplitude is identical to the result in 
equation~\ref{eqn:amp}. Likewise, the mean amplitude of the measured vector is given 
by equation~\ref{eqn:avgamp}. It follows that one cannot distinguish between 
linear and circular polarization on the basis of the amplitude distribution alone.

  The joint probability density given by equation~\ref{eqn:jdc} is independent of and 
symmetric in azimuth. Therefore, the orientation of the polarization vector varies 
primarily in colatitude, $\theta$, and is distributed according to
\begin{equation}
f(\theta)  =  \int_{0}^{2\pi}\int_{0}^{\infty}f(r,\theta,\phi)dr d\phi,
\end{equation}

\begin{eqnarray}
f(\theta) & = & {\sin\theta\over{2}}\Biggl\{\exp{\Biggl(-{s^2\sin^2\theta\over {2}}
          \Biggr)}\Biggl[1+{\rm erf}\Biggl({s\cos\theta\over{\sqrt{2}}}\Biggr)\Biggr]
            (1 + s^2\cos^2\theta) 
          \nonumber \\
          & + & \sqrt{{2\over{\pi}}}s\cos\theta
            \exp{\Biggl(-{s^2\over{2}}\Biggr)}\Biggr\}.
\label{eqn:colat}
\end{eqnarray}
The colatitude distribution has only one free parameter, $s$, and is shown in the 
bottom panel of Figure~\ref{fig:fixed}. When $s$ is large ($s>>1$), the colatitude 
distribution approaches the Rayleigh distribution.

\begin{equation}
f(\theta) \simeq s^2\theta\exp{\Biggl(-{s^2\theta^2\over {2}}\Biggr)}
\end{equation}

\begin{figure}
\plotone{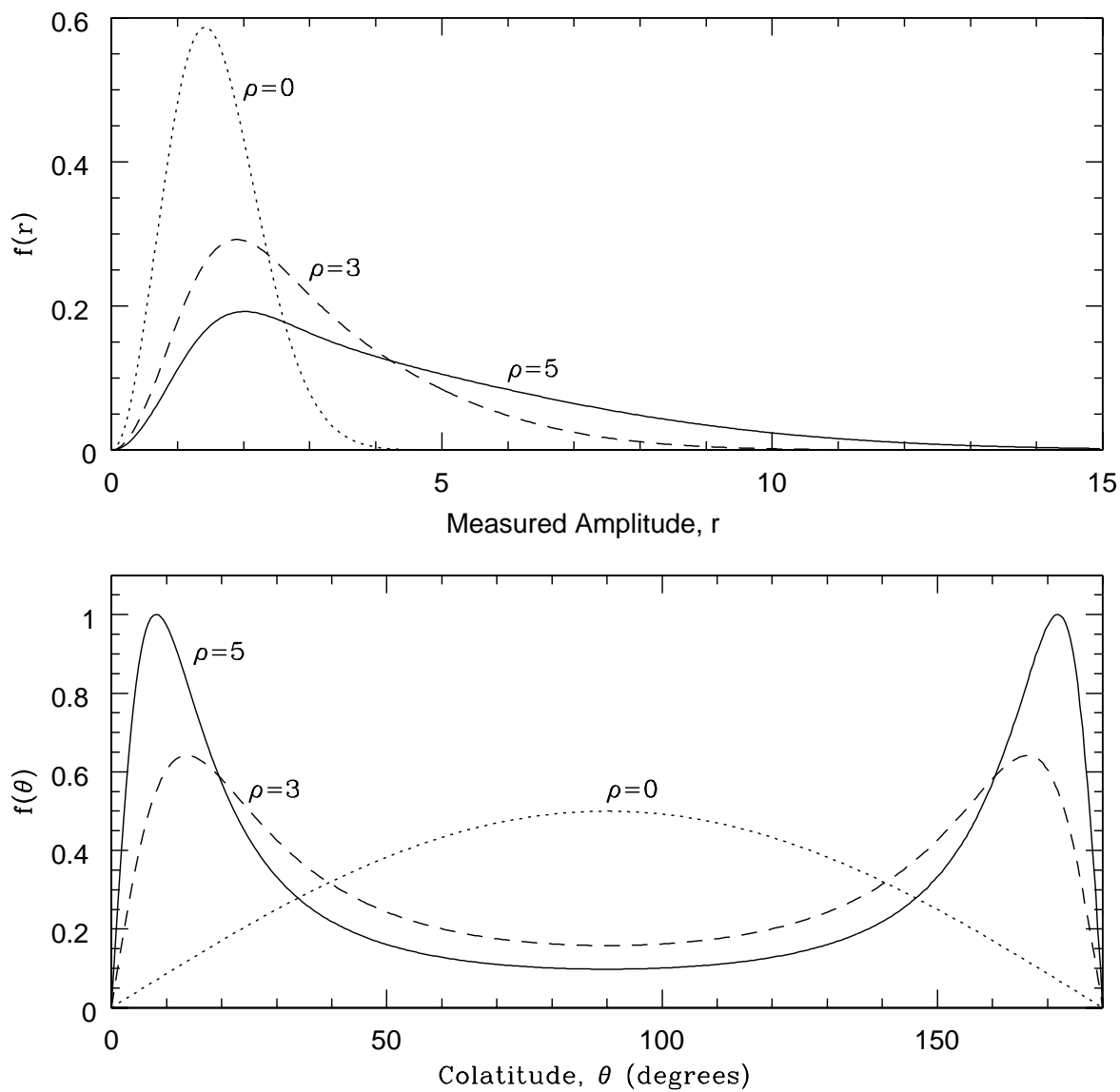}
\caption{Distributions of a polarization vector's measured amplitude (top panel)
and colatitude (bottom panel) when circularly-polarized OPM occurs in the radiation 
and the mean intrinsic polarization is $p=0$. In this case, the radiation's 
polarization fluctuates, and the different curves represent the distributions 
expected for different mode fluctuation ratios, $\rho$. The value of the 
instrumental noise used in the amplitude distributions is $\sigma_N=1$.}
\label{fig:opm}
\end{figure}

\section{ORTHOGONAL MODES OF POLARIZATION}
\label{sec:opm}

  Orthogonal polarization modes are commonly observed in the radio emission
from pulsars (e.g. Manchester, Taylor, \& Huegunin 1975; Cordes, Rankin, \& Backer 
1978; Backer \& Rankin 1980; Stinebring et al. 1984). In a rather confusing use of 
the term, the fact that the modes are {\it orthogonal} actually means their 
polarization vectors are {\it antiparallel} in Poincar\'e space. The modes also 
appear to occur at the same time (McKinnon \& Stinebring 1998, hereafter MS1; 
McKinnon \& Stinebring 2000, hereafter MS2). So although the individual modes may 
be highly polarized, their simultaneous interaction can cause the observed 
polarization to be small. To account for the statistical fluctuations observed in 
the polarization of pulsar radio emission, the amplitudes of the mode polarization 
vectors can be represented by the random variables $X_1$ and $X_2$ (MS1; MS2; 
McKinnon 2002). The means of these random variables shall be designated as $\mu_1$ 
and $\mu_2$, and their standard deviations shall be denoted as $\sigma_1$ and 
$\sigma_2$. Since the observations indicate that the modes occur simultaneously and 
their polarization vectors are antiparallel, the intrinsic polarization fluctuations 
occur along a diagonal in Poincar\'e space. For a diagonal specified by an azimuth 
$\phi_o$ and a colatitude $\theta_o$, the observed Stokes parameters can be 
described by (MS2; McKinnon 2002)
\begin{equation}
{\rm Q} = \sin\theta_o\cos\phi_o (X_1 - X_2) + X_{\rm N,Q},
\end{equation}
\begin{equation}
{\rm U} = \sin\theta_o\sin\phi_o (X_1 - X_2) + X_{\rm N,U},
\end{equation}
\begin{equation}
{\rm V} = \cos\theta_o (X_1 - X_2) + X_{\rm N,V}.
\end{equation}
  
  The equations presented so far are valid regardless of the type of random 
variable that actually describes the amplitudes of the mode polarization vectors. 
The probability density of these random variables must be known before the joint 
probability density can be derived. In the examples which follow, the mode 
polarization amplitudes are assumed to be independent, Gaussian random variables.

\subsection{Linearly-Polarized Modes}

  Consider the case when the modes are linearly polarized ($\theta_o = \pi/2$) 
and their polarization fluctuations occur in Stokes Q ($\phi_o = 0$). In this 
case, the fluctuations in the measured values of U and V are due to the 
instrumental noise. Therefore, they are Gaussian random variables with zero means 
($\mu_U=\mu_V=0$) and standard deviations of $\sigma_U=\sigma_V=\sigma_N$. Since 
the mode polarization amplitudes are assumed to be Gaussian random variables, the 
measured Stokes Q is also a Gaussian random variable with mean 
$\mu_Q=p=\mu_1-\mu_2$ and a standard deviation of 
$\sigma_Q=(\sigma_1^2 + \sigma_2^2 + \sigma_N^2)^{1/2}\equiv\sigma_N(1+\rho^2)^{1/2}$.
The quantity $\rho$ is a measure of the intrinsic polarization fluctuations relative 
to the instrumental noise. With these assumptions, the modes will occur at PAs
of $\psi=0^\circ$ and $\psi=90^\circ$, and will occur with equal frequency if the 
mode polarizations are the same on average (i.e. $\mu_1=\mu_2$; $\mu_Q=0$). The 
measured Stokes parameters are again independent of one another, and their joint 
probability density is the product of their individual probability densities
(see, also, Appendix A3 of MS1). When the modes have identical mean values 
($\mu_1=\mu_2$; $\mu_Q=0$), the joint probability density is

\begin{equation}
f(r,\theta,\phi) = {r^2\over{\sigma_N^3}}{\sin\theta\over{(2\pi)^{3/2}}}
                   {1\over {(1+\rho^2)^{1/2}}}\exp\Biggl[-{r^2(1+\rho^2)
             - r^2\rho^2\cos^2\phi\sin^2\theta\over{2(1+\rho^2)\sigma_N^2}}\Biggr],
\end{equation}

\noindent and the distribution of the measured polarization vector's orientation is

\begin{equation}
f(\theta,\phi) = \int_{0}^{\infty} f(r,\theta,\phi)dr 
               = {\sin\theta\over{4\pi}}{1 + \rho^2\over{[1+\rho^2
                 (1-\cos^2\phi\sin^2\theta)]^{3/2}}}.
\label{eqn:Lorient}
\end{equation}

\subsection{Circularly-Polarized Modes}

  Although the observations indicate that OPM is predominantly linearly polarized,
consider an idealized case where OPM is circularly polarized. The assumption of 
circularly polarized modes, combined with the observational facts that the modes 
occur simultaneously and their polarization vectors are antiparallel, means that 
the intrinsic polarization fluctuations occur only in the Stokes parameter V
($\theta_o = 0$). Again assuming that the mode polarizations are Gaussian random 
variables with identical mean values, the joint probability density of the measured 
polarization vector's amplitude and orientation is

\begin{equation}
f(r,\theta,\phi) = {r^2\over{\sigma_N^3}}{\sin\theta\over{(2\pi)^{3/2}}}
                   {1\over {(1+\rho^2)^{1/2}}}\exp\Biggl[-{r^2(1+\rho^2\sin^2\theta)
                   \over{2(1+\rho^2)\sigma_N^2}}\Biggr].
\label{eqn:jdcm}
\end{equation}

The assumption of circularly polarized modes introduces azimuthal symmetry to the 
3D statistics, and allows for a straightforward, analytical derivation of the 
amplitude distribution.

\begin{equation}
f(r) = {r^2\over{\sigma_N^3}}\sqrt{{2\over{\pi}(1+\rho^2)}}
       \exp\Biggl(-{r^2\over {2\sigma_N^2}}\Biggr)
       \int_0^1\exp\Biggl[{\rho^2r^2u^2\over{2(1+\rho^2)\sigma_N^2}}\Biggr]du
\label{eqn:ampcm}
\end{equation}

\noindent The amplitude distribution is shown in the top panel of 
Figure~\ref{fig:opm} for different values of $\rho$. The figure shows that
the modes tend to stretch what was originally the Maxwell-Boltzmann distribution 
of the instrumental noise to a highly asymmetric distribution. The extent of the 
tail on this distribution increases with the magnitude of the mode fluctuations,
$\rho$.

  The azimuthal symmetry of equation~\ref{eqn:jdcm} also means the orientation 
of the polarization vector varies primarily in colatitude.

\begin{equation}
f(\theta) = {\sin\theta\over{2}}{1 + \rho^2\over{(1+\rho^2\sin^2\theta)^{3/2}}}
\label{eqn:Vorient}
\end{equation}

\noindent The colatitude distribution is shown in the bottom panel of 
Figure~\ref{fig:opm} for different values of $\rho$. The distribution is 
isotropic for $\rho=0$, and the polarization modes become more obvious as
$\rho$ increases. The two peaks in the colatitude distribution do not occur 
precisely at $\theta=0, \pi$ because the instrumental noise in Stokes Q and U 
weights the orientation of the polarization vector towards $\theta=\pi/2$.

  Equations~\ref{eqn:Lorient} and~\ref{eqn:Vorient} were derived for polarization
fluctuations along specific diagonals in Poincar\'e space. For fluctuations
along any other diagonal specified by a colatitude $\theta_o$ and an azimuth 
$\phi_o$, the possible orientations of the measured polarization vector are 
described by
\begin{equation}
f(\theta,\phi) = {\sin\theta\over{4\pi}}{1 + \rho^2\over{[1+\rho^2g(\theta,
                  \phi)]^{3/2}}},
\end{equation}
where 
\begin{equation}
g(\theta,\phi) = 1-[\cos\theta_o^2\cos^2\theta + \sin^2\theta_o\sin^2\theta
                  (\cos^2\phi\cos^2\phi_o + \sin^2\phi\sin^2\phi_o)].
\end{equation}

\section{DIRECTIONAL STATISTICS}
\label{sec:direct}

  The directional statistics of a polarization vector predict the possible
orientations of the vector at a given vector amplitude, $r_o$. They are
determined from the 2D or 3D statistics by computing the conditional density 
of the vector's orientation. In the specific 2D case when only the linear 
polarization is considered, the conditional density gives the possible PA values 
on a circle of radius $r_o$. In the more general 3D case, the conditional density 
gives the possible vector orientations on a sphere of radius $r_o$.

\subsection{Distribution on a Circle}
\label{sec:circle}

\begin{figure}
\plotone{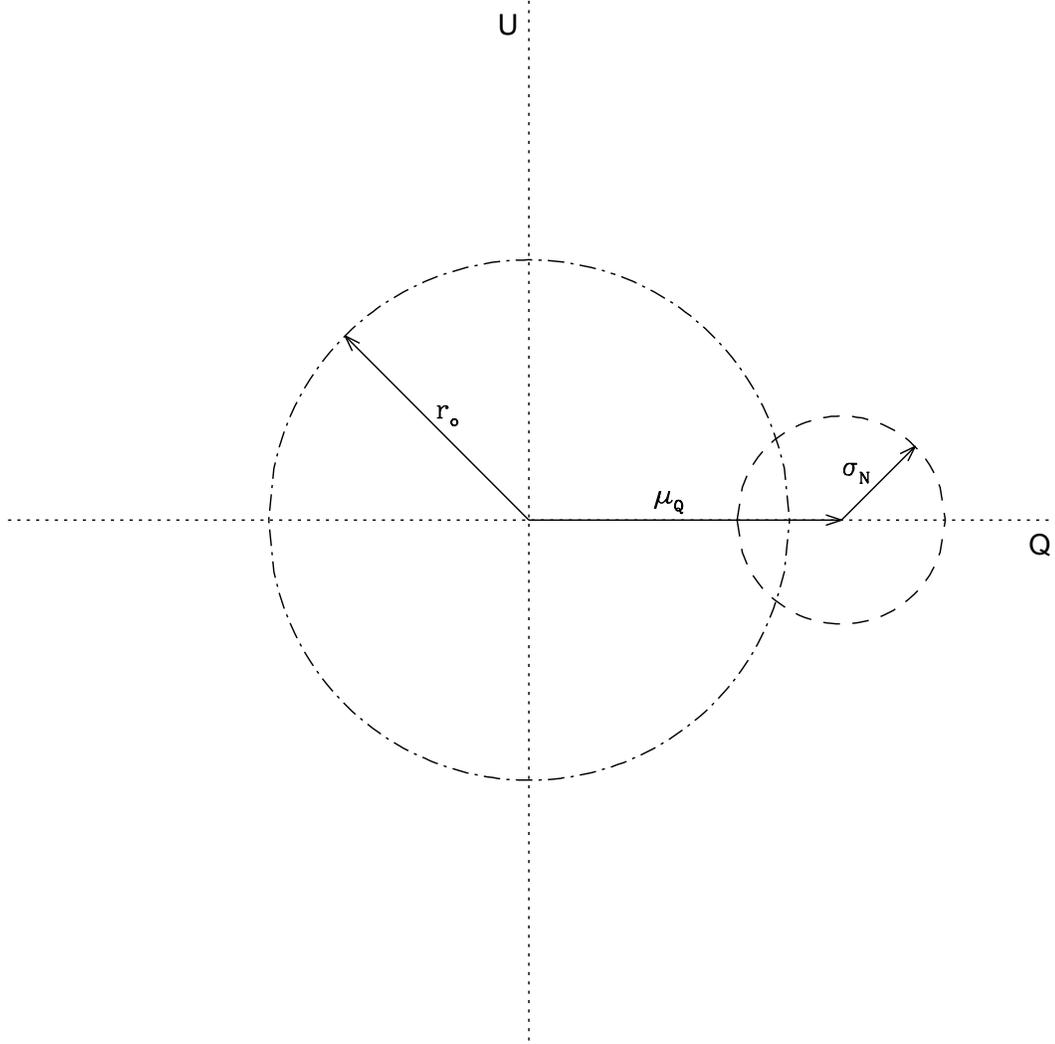}
\caption{Graphical representation of the calculation of the PA conditional density. 
The radiation's polarization vector has an amplitude $\mu_Q$ and points in the 
direction of increasing Q in the Q-U plane of Poincar\'e space. The measured
polarization vector includes the instrumental noise, which is a vector with an
amplitude characterized by $\sigma_N$ that can point in any direction (dashed circle). 
The PA conditional density is the distribution of measured PAs along the circle of 
radius $r_o$.}
\label{fig:circle}
\end{figure}

  Figure~\ref{fig:circle} shows a linear polarization vector in the Q-U plane of 
Poincar\'e space, and illustrates the calculation of the PA conditional density.
The measured vector is the vector sum of the radiation's polarization and the 
randomly polarized instrumental noise. The conditional density is the distribution 
of the measured vector's PAs along the circle of radius, $r_o$, and is given by

\begin{equation}
f_{\psi | r_o}(\psi) = {f(r_o,\psi)\over{f(r_o)}}.
\label{eqn:cond_dens}
\end{equation}

\noindent For the geometry displayed in the figure and noting that $\psi$ is 
defined on the interval $-\pi/2 \le\psi\le\pi/2$, the joint probability density 
and the amplitude distribution are, respectively 
\begin{equation}
f(r,\psi) = {r\over{\sigma_N^2}}{1\over{\pi}}
          \exp{\Biggl[-{(r^2+\mu_Q^2-2\mu_Qr\cos{2\psi})\over{2\sigma_N^2}}\Biggr]},
\end{equation}
\begin{equation}
f(r)={r\over{\sigma_N^2}}\exp\Biggl[-{(r^2+\mu_Q^2)\over{2\sigma_N^2}}\Biggr]
     I_0\Biggl({r\mu_Q\over{\sigma_N^2}}\Biggr),
\label{eqn:vonmises}
\end{equation}

\noindent where $I_0(x)$ is the modified Bessel function of order zero. From 
Equations~\ref{eqn:cond_dens} through~\ref{eqn:vonmises}, one can show that the 
PA conditional density is a von Mises distribution (Mardia 1972; see also MS1) 
with a dispersion parameter of $b=r_o\mu_{\rm Q}/\sigma_{\rm N}^2$. 
\begin{equation}
f_{\psi | r_o}(\psi) =
     {\exp (b\cos{2\psi})\over {\pi I_0(b)}}, \qquad -\pi/2\le\psi\le\pi/2.
\end{equation}

\begin{figure}
\plotone{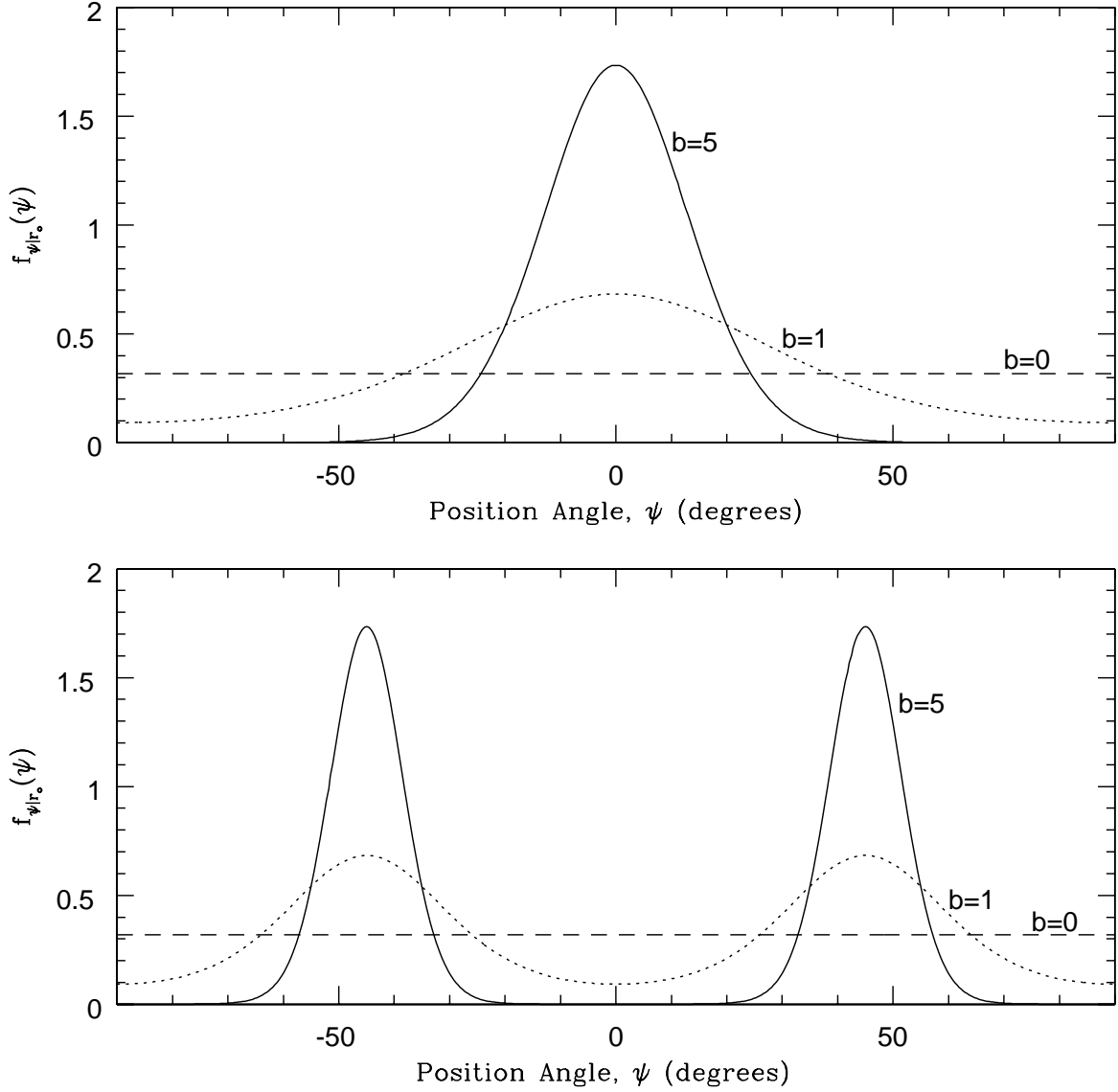}
\caption{Conditional density for the position angle (PA) of a linear polarization
vector. When the radiation's polarization is fixed, the conditional density
follows the von Mises distribution (top panel). For linearly-polarized OPM,
the conditional density is a bimodal von Mises distribution (bottom panel).
The distributions are shown for different dispersion parameters, $b$.}
\label{fig:vm}
\end{figure}

\noindent The PA conditional density is shown in the top panel of Figure~\ref{fig:vm} 
for different values of $b$. It is a uniform distribution at $b=0$, and is
a Gaussian distribution with a FWHM of $(2\ln 2/b)^{1/2}$ when $b$ is large .

  For linearly-polarized OPM, the joint probability density is (MS1)
\begin{equation}
f(r,\psi) = {r\over{\sigma_N^2(1+\rho^2)^{1/2}}}{1\over{\pi}}
             \exp{\Biggl[-{r^2(1+\rho^2\sin^2{2\psi})+\mu_Q^2-2\mu_Qr\cos{2\psi}
             \over{2\sigma_N^2(1+\rho^2)}}\Biggr]}
\end{equation}
where $\rho=(\sigma_1^2+\sigma_2^2)^{1/2}/\sigma_{\rm N}$. When the modes have the 
same polarization on average ($\mu_Q=0$), the amplitude distribution is

\begin{equation}
f(r)={r\over{\sigma_N^2(1+\rho^2)^{1/2}}}\exp\Biggl[-{r^2\over{4\sigma_N^2}}
     {(2+\rho^2)\over{(1+\rho^2)}}\Biggr] I_0\Biggl[{r^2\over{4\sigma_N^2}}
     {\rho^2\over{(1+\rho^2)}}\Biggr].
\end{equation}

\noindent One can then show that the PA conditional density for linearly-polarized 
OPM is a bimodal von Mises distribution with a dispersion parameter of 
$b=r_o^2\rho^2/4\sigma_N^2(1+\rho^2)$.
\begin{equation}
f_{\psi | r_o}(\psi) = {\exp [b\cos(4\psi)]\over {\pi I_0(b)}}, 
                     \qquad -\pi/2\le\psi\le\pi/2
\label{eqn:bvm}
\end{equation}

\noindent The bimodal von Mises distribution is shown in the bottom panel of 
Figure~\ref{fig:vm} for different values of $\rho$. For display purposes, 
equation~\ref{eqn:bvm} was modified by replacing $\psi$ with $\psi-\pi/4$ so 
that the modes occur at $\psi=-\pi/4, \pi/4$ in the figure.

  Detection thresholds are commonly applied to the measured amplitude of the linear 
polarization vector before constructing PA histograms (e.g. Stinebring et al. 
1984; MS1). The resulting histogram is not an exact replica of the underlying PA 
distribution, but is more similar to a PA conditional density. 

\subsection{Distribution on a Sphere}

  When all three Stokes parameters are measured, the conditional density of
a polarization vector's orientation is the distribution of vector orientations 
on the surface of a sphere, instead of the perimeter of a circle as in 
\S~\ref{sec:circle}. As with the PA conditional density, the conditional density 
of the vector's orientation is calculated from the joint probability density and 
the amplitude distribution. 

\begin{figure}
\plotone{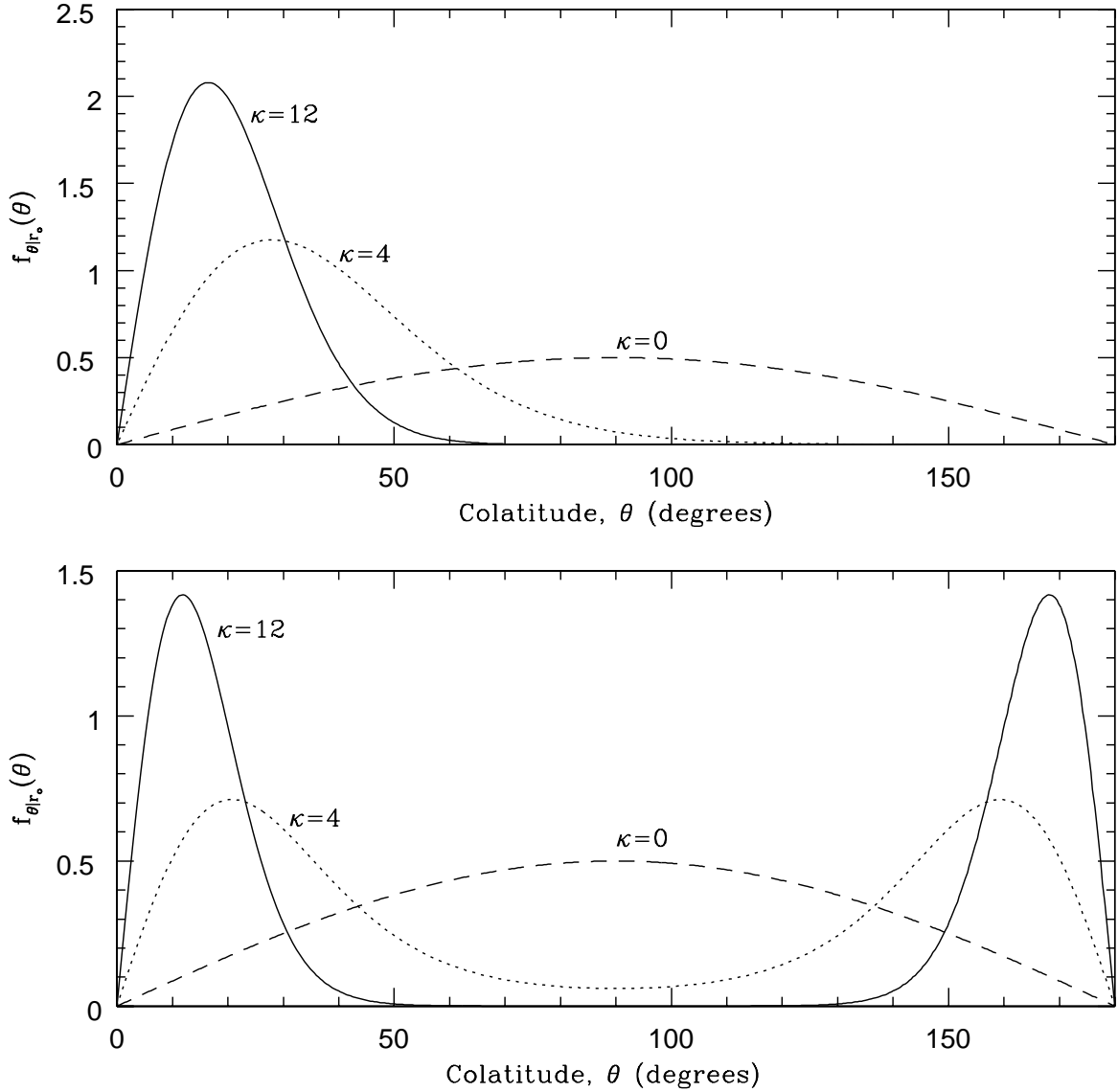}
\caption{Conditional density of a polarization vector's colatitude, $\theta$, when 
the radiation is circularly polarized. When the radiation's polarization is fixed, 
the conditional density follows the Fisher distribution (top panel). In the case 
of circularly-polarized OPM, the conditional density follows a Watson bipolar
distribution (bottom panel). The distributions are shown for different concentration
parameters, $\kappa$.}
\label{fig:fiswat}
\end{figure}

  For a radiation source with fixed polarization properties, the joint probability
density is given by equation~\ref{eqn:jd} and the amplitude distribution is given 
by equation~\ref{eqn:amp}. Since the distribution of the polarization vector's 
amplitude is independent of its orientation as discussed in \S~\ref{sec:fixed}, 
one can use the two equations to show that the conditional density always follows 
the general Fisher distribution,
\begin{equation}
f(\theta,\phi |r_o)
             = {f(r_o,\theta,\phi)\over{f(r_o)}}
             = {\kappa\sin\theta\over{4\pi\sinh(\kappa)}}
                \exp\{\kappa[\cos\theta_o\cos\theta + \sin\theta_o\sin\theta
                \cos(\phi-\phi_o)]\},
\label{eqn:fisher}
\end{equation}

\noindent where $\kappa=pr_o/\sigma_N^2$ is a concentration parameter (Fisher, 
Lewis, \& Embleton 1992). As a specific example, $\theta_o=0$ when the radiation is 
circularly polarized, and from Equation~\ref{eqn:fisher} the conditional density is
\begin{equation}
f(\theta,\phi |r_o)
             = {\kappa\sin\theta\over{4\pi\sinh(\kappa)}}\exp(\kappa\cos\theta).
\label{eqn:circfish}
\end{equation}
Equation~\ref{eqn:circfish} is independent of azimuth, as one would expect from the 
azimuthal symmetry of the problem. Therefore, integration of $f(\theta, \phi | r_0)$
over $\phi$ yields a factor of $2\pi$, and the resulting colatitude conditional 
density is $f_{\theta |r_o}(\theta)= 2\pi f(\theta, \phi | r_0)$. The colatitude
conditional density is shown in the top panel of Figure~\ref{fig:fiswat} for 
different values of $\kappa$. Equation~\ref{eqn:circfish} can also be derived 
directly from the joint probability density in Equation~\ref{eqn:jdc} and the 
amplitude distribution in Equation~\ref{eqn:amp}. 

  For the specific case of circularly-polarized OPM, the conditional density 
can be found from Equations~\ref{eqn:jdcm} and~\ref{eqn:ampcm}. The conditional 
density is a Watson bipolar distribution (Fisher, Lewis, \& Embleton 1992) with 
a concentration parameter of $\kappa=r_o^2\rho^2/2(1+\rho^2)\sigma_N^2$,
\begin{equation}
f(\theta,\phi |r_o)={\sin\theta\over{4\pi w(\kappa)}}\exp(\kappa\cos^2\theta).
\label{eqn:watson}
\end{equation}
Here, $w(\kappa)$ is a normalization factor given by 
\begin{equation}
w(\kappa)=\int_0^1\exp(\kappa u^2)du.
\end{equation}
This particular distribution is also independent of azimuth, so the colatitude
conditional density is again $f_{\theta |r_o}(\theta)=2\pi f(\theta,\phi |r_o)$. 
The colatitude conditional density is shown in the bottom panel of 
Figure~\ref{fig:fiswat} for different concentration parameters. Both the Fisher 
and Watson bipolar distributions become isotropic distributions in the limit 
$\kappa=0$. 

\section{DISCUSSION}
\label{sec:discuss}

  The essential point of this paper is microwave polarization measurements can be
interpreted in the context of a joint probability density of the Stokes parameters
Q, U, and V because the large photon densities at radio wavelengths allow their
simultaneous detection. These same photon statistics make coherent phase detection
possible and thus the measurement of visibility phase for radio aperture synthesis
observations (Radhakrishnan 1999). This important realization allows the derivation
of the general 3D statistics for a measured polarization vector. The 3D statistics, 
in turn, can be used to extract information about the radiation's polarization 
properties.  The statistics derived here are unique to radio polarization 
measurements, particularly when the radiation's polarization fluctuates on a 
timescale comparable to the sampling interval. The statistics may not apply to 
polarization observations at much shorter wavelengths because the Stokes parameters 
cannot be measured simultaneously and the polarization state of the radiation may 
change between consecutive measurements.

  The 3D statistics provide the appropriate framework for interpreting observations 
where both the linear and circular polarization are significant and/or highly 
variable, as is often the case in pulsar radio emission. Historically, but with 
some notable exceptions (e.g. Manchester, Taylor, \& Huguenin 1975), the linear 
and circular polarization of individual pulses have been interpreted or evaluated 
separately, making it difficult to visualize or determine how the polarization 
vector's orientation varies from pulse-to-pulse. By applying the 3D statistics 
derived in this paper, one may be able to make more definite statements about the 
true polarization state of the radiation.

  The statistical model used to derive the polarization statistics accurately 
describes most radio polarization observations. The model is a set of linear 
equations that accounts for the radiation's polarization properties and includes 
the instrumental noise. The model is valid as long as the telescope system 
temperature, $T_S$, is greater than the antenna temperature produced by the radio 
source, $T_A$. When $T_A$ becomes comparable with $T_S$, the instrumental noise is
not independent between Stokes parameters and the model is no longer linear. 

  In deriving the 3D statistics, special emphasis was placed on the specific 
case of circularly-polarized radiation. This apparent bias does not signify a 
shortcoming of the statistical model's ability to describe other types of 
polarization, but merely highlights the fact that straightforward, analytical 
solutions can be derived given the azimuthal symmetry provided by the geometry
(e.g. Eqns.~\ref{eqn:colat} and~\ref{eqn:watson}). Also, the functional form of 
the distribution of the polarization vector's orientation is easier to illustrate 
(e.g. Figs.~\ref{fig:fixed}, \ref{fig:opm}, and~\ref{fig:fiswat}) for 
circularly-polarized radiation because the distribution contains only one 
independent variable (colatitude).

  Both the 3D and directional statistics for microwave polarimetry are remarkably 
similar to those that are relevant in other fields of the physical, biological, 
and Earth sciences, and it is astonishing that such diverse measurements can be 
described by the same statistics. For example, measurements of remanent 
magnetization in rock formations are consistent with the Fisher distribution, 
and measurements of the intersections between the cleavage and bedding planes of 
certain types of minerals follow the Watson bipolar distribution (Fisher, Lewis, 
\& Embleton 1987, and references therein). The von Mises distribution can describe 
the orientations of sand-grains and the vanishing angles of birds taking flight 
(Mardia 1972, and references therein). These similarities hold great promise for 
future polarization work because the mature, rigorous, statistical methods 
developed in other fields may be applied to radio polarization observations.

\end{document}